\documentclass[conference]{IEEEtran}

\IEEEoverridecommandlockouts
\usepackage{cite}
\usepackage{amsmath,amssymb,amsfonts}
\usepackage{algorithm}
\usepackage{algorithmic}

\usepackage{graphicx}
\usepackage{fancyhdr}
\usepackage{pifont}
\usepackage{wasysym}
\usepackage{utfsym}
\usepackage{enumerate}
\usepackage{textcomp}
\usepackage{listings}
\usepackage{xcolor}
\usepackage{tcolorbox}
\usepackage{longtable}
\usepackage{caption}
\usepackage{subcaption}
\usepackage{booktabs} 
\usepackage{colortbl}
\makeatletter
\newcommand{\rmnum}[1]{\romannumeral #1}
\newcommand{\Rmnum}[1]{\expandafter\@slowromancap\romannumeral #1@}
\makeatother
\usepackage[font=small]{caption}
\usepackage{multirow}
\usepackage{multicol}
\usepackage{bm}
\usepackage[numbers,sort&compress]{natbib} 
\usepackage{lipsum}
\def\BibTeX{{\rm B\kern-.05em{\sc i\kern-.025em b}\kern-.08em
    T\kern-.1667em\lower.7ex\hbox{E}\kern-.125emX}}
    
\fancypagestyle{firstpagefooter}{%
  \fancyhf{}
  
}

\usepackage{footmisc} 

\definecolor{pink}{rgb}{1, 0.5, 0.5}
\definecolor{mygreen}{RGB}{0, 176, 80} 
\definecolor{myred}{RGB}{222, 26, 55}
\definecolor{myyellow}{RGB}{255, 152, 0}
\definecolor{myblue}{RGB}{0, 122, 192}
\lstdefinestyle{mystyle}{
	basicstyle=\ttfamily\scriptsize\bfseries,
	frame=single,
	framerule=0.7pt,
	breaklines=true,
	numberstyle=\tiny\color{gray},
	numbersep=8pt,
	keywordstyle=\color{blue}\bfseries,
	commentstyle=\color{green},
	stringstyle=\color{orange},
	showstringspaces=false,
	escapeinside={(*@}{@*)},	
}

\lstdefinestyle{mystyle1}{
	keywords={module, input, output, reg, wire, always, posedge, negedge, if, else, begin, end, endmodule},
	keywordstyle=\color{blue}\bfseries,
	morecomment=[l]{//},
	commentstyle=\color{green},
	basicstyle=\ttfamily\footnotesize\bfseries,  
	numbers=left,                       
	numberstyle=\tiny\color{gray},      
	numbersep=8pt,                      
	frame=single,   
	framerule=0.7pt,                      
	showstringspaces=false,             
	breaklines=true,                    
	escapeinside={(*@}{@*)},
}

\begin{document}

\title{LintLLM: An Open-Source Verilog Linting Framework Based on Large Language Models\vspace{-0.5em}}

\author{
	\IEEEauthorblockN{Zhigang Fang\textsuperscript{1}, Renzhi Chen\textsuperscript{2}*, Zhijie Yang\textsuperscript{3}, Yang Guo\textsuperscript{1}, Huadong Dai\textsuperscript{3}, Lei Wang\textsuperscript{3}*}
	\IEEEauthorblockA{\textsuperscript{1}\textit{National University of Defense Technology}, Changsha, China \\
		\textsuperscript{2}\textit{Qiyuan Lab}, Beijing, China \\
		\textsuperscript{3}\textit{Defense Innovation Institute, Academy of Military Sciences}, Beijing, China \\
		Email: $\{$fangzhigang, leiwang$\}$@nudt.edu.cn, chenrenzhi@qiyuanlab.com}\vspace{-3em}
}

\maketitle

\renewcommand{\footnoterule}{%
	\hrule width 0.4\textwidth height 0.5pt \kern 2pt 
}

\makeatletter
\renewcommand{\@makefntext}[1]{%
	\leftskip=0pt 
	\noindent 
	\@makefnmark\hspace{0.5em}#1%
}

\footnotetext {*Co-Corresponding author: Renzhi Chen and Lei Wang. \vspace{-2.5em}}

\begin{abstract}

Code Linting tools are vital for detecting potential defects in Verilog code. However, the limitations of traditional Linting tools are evident in frequent false positives and redundant defect reports. Recent advancements in large language models (LLM) have introduced new possibilities in this area. In this paper, we propose LintLLM, an open-source Linting framework that utilizes LLMs to detect defects in Verilog code via Prompt of Logic-Tree and Defect Tracker. Furthermore, we create an open-source benchmark using the mutation-based defect injection technique to evaluate LLM's ability in detecting Verilog defects. Experimental results show that o1-mini improves the correct rate by 18.89\% and reduces the false-positive rate by 15.56\% compared with the best-performing EDA tool. Simultaneously, LintLLM operates at less than one-tenth of the cost of commercial EDA tools. This study demonstrates the potential of LLM as an efficient and cost-effective Linting tool for hardware design. The benchmark and experimental results are open-source at URL: https://github.com/fangzhigang32/Static-Verilog-Analysis

\end{abstract}


\thispagestyle{firstpagefooter}

\section{Introduction}

Ensuring code quality and maintaining consistent coding styles is essential for producing robust and defect-free Register-Transfer Level (RTL) designs \cite{deng2023verilog}. Code Linting tools can significantly reduce verification costs by analyzing source code for potential defects, such as inconsistent coding styles and unsynthesizable constructs, which differ from bugs as they do not necessarily lead to errors \cite{stefanovic2020static}. Commercial EDA tools (e.g. SpyGlass \cite{SpyGlass}) detect Verilog defects via matching Design Under Test (DUT) with predefined rules, which pattern causes a large number of false positives \cite{novak2010taxonomy}. Although customizing rules that meet the design specifications can alleviate this symptom, analyzing and extracting these rules is difficult. Additionally, expensive licensing fees burden small and medium-sized enterprises and research institutions. Previous open-source EDA tools (e.g. Verilator \cite{Verilator}) tended to exhibit lower performance, which increases the complexity for designers to deal with code defects and prolongs debugging time \cite{firdous2019speeding}. These challenges affect the promotion and popularity of traditional Linting tools, highlighting the need for a comprehensive and open-source Linting tool.

Recently, LLMs such as GPT-4 have demonstrated significant potential in Verilog code debugging. In terms of function bugs, LLMs combines design specifications and DUTs to achieve efficient bug localization without testbench \cite{journals/corr/abs-2409-15186}. For syntax bugs, RTLFixer \cite{conf/dac/TsaiLR24} utilizes the Retrieval Augmented Generation (RAG) strategy to greatly improve the accuracy and efficiency of automatically repairing bugs. To optimize the debugging process, MEIC \cite{journals/corr/abs-2405-06840} proposes an iterative framework to accelerate the debugging of complex hardware through multiple rounds of feedback. LLM4SecHW \cite{conf/asianhost/FuYDGQ23} and HDLdebugger \cite{journals/corr/abs-2403-11671} further improve the LLM's application effect in hardware debugging through domain data fine-tuning and reverse engineering. These studies highlight the vast potential of LLMs in assisting RTL Coding and opening new avenues for EDA. However, most existing research primarily focuses on the application of LLMs in assisting code debugging. In contrast, our study explores the capabilities of LLMs for static code analysis. This work further promotes the refined application of LLMs in chip design automation. 



In this paper, we develop LintLLM, an open-source Linting framework that leverages LLM to detect defects in Verilog code, which finds defects earlier than the compiler. It enhances the capability of LLMs via the \emph{Prompt of Logic-Tree} and \emph{Defect Tracker} we proposed, which address the limitations of traditional Linting tools that exhibit redundancy and false positives. To evaluate the effectiveness of LintLLM, we create a high-quality benchmark consisting of 90 Verilog designs, encompassing 11 types of common code defects.


We evaluated the code defect detection capabilities of five LLMs and two traditional Linting tools on our benchmark. Experimental results show that o1-mini improves the correct rate by 18.89\% and reduces the false-positive rate by 15.56\% compared with the best-performing EDA tool. Moreover, LintLLM operates at less than one-tenth of the cost of the commercial EDA tool. This study demonstrates the potential of LLM as an effective Linting tool for hardware design and provides a cost-effective alternative for detecting code defects.

Our main contributions are as follows:
\begin{itemize}

	\item 
	We proposed a Linting framework named LintLLM that leverages LLMs to detect code defects. To the best of our knowledge, this is the first work employing LLM as a novel Linting tool in the hardware design domain.
	
	\item 
	We created an open-source benchmark to evaluate LLM's ability for detecting Verilog defects. It covers 11 defect types and includes 90 Verilog designs.
	
	\item 
	We introduced a template of Prompt Engineering named Prompt of Logic-Tree. It can translate complex algorithms into tree-structured prompts that are easier for LLMs to understand.
	
	\item 
	We proposed the Defect Tracker. It can locate the main defect causing multiple secondary defects in the code, reducing the redundancy of defect reports.
	
\end{itemize}



\section{Background and Related Works}

\subsection{LLMs for Chip Design}

The chip design process includes system specification, architectural design, functional design, logic synthesis, and physical design \cite{conf/aspdac/LiHT0ZWLQLLSCBZ24}. Recent advancements highlight the application of LLMs in chip design across four primary directions, RTL code generation, EDA script generation, chip verification, and domain-specific knowledge Q\&A. In RTL code generation, researchers have employed LLMs to generate Verilog code through prompt engineering techniques \cite{journals/corr/abs-2305-14019,journals/corr/abs-2311-04887}. Other studies have focused on instruction fine-tuning to adapt LLMs specifically for Verilog code generation, significantly improving the quality and accuracy of generated code \cite{journals/todaes/ThakurAPTDKG24,conf/icml/PeiZYH024,journals/corr/abs-2407-10424,journals/corr/abs-2407-18333}. Additionally, research has demonstrated the use of LLM-based agents that iteratively improve code quality by integrating feedback from EDA tools \cite{journals/corr/abs-2312-08617,journals/corr/abs-2408-08927}. In script generation, LLMs have been utilized to automatically create scripts for EDA tools operations and simplify workflows \cite{conf/dac/ChangWY0JZCLYZZ24,journals/tcad/WuHZYZZY24}. To evaluate the syntax and function correctness of generated codes, researchers have developed open-source datasets \cite{conf/aspdac/LuLZX24,journals/corr/abs-2407-01910}. However, these datasets currently lack mechanisms to evaluate defect detection capabilities. For chip verification, LLMs have also been employed to generate test stimuli and analyze coverage, demonstrating their utility in comprehensive testing tasks \cite{journals/corr/abs-2406-04373,journals/corr/abs-2405-02326,conf/date/XiaoDYCWZDWTX24,journals/corr/abs-2310-04535}. Moreover, researchers have explored LLM as an interactive design assistant to assist preliminary design stages through iterative dialogue with designers \cite{conf/mlcad/BlockloveGKP23,journals/corr/abs-2311-00176}. 

Overall, these studies underscore the robust capabilities of LLMs in supporting EDA tools across various chip design tasks. In contrast to prior studies, our research investigates the potential of LLM-native approaches \cite{chen2024large}, employing LLM as an open-source Linting tool to autonomously detect Verilog code defects without relying on EDA tools supporting for the first time.

\subsection{Static Code Analysis}


Static code analysis is a method of detecting syntax, coding style, and design inconsistencies in code through predefined patterns or rules without actually running the code \cite{journals/ese/VassalloPPPGZ20}. Linting tools can assist verification personnel in improving defect detection efficiency. They work by analyzing the syntax, structure, and potential defects of the code and identifying potential problems that do not comply with these rules by matching them with the established rule set \cite{novak2010taxonomy}. In the field of hardware design, traditional EDA tools such as Synopsys's SpyGlass \cite{SpyGlass}, Cadence's JasperGold \cite{JasperGold}, and Siemens's Questa Lint \cite{QuestaLint} can detect potential defects in Verilog through customized coding rules. However, a common problem with these tools is that they tend to generate false positive reports, reporting code issues that actually are not defects \cite{conf/icse/JohnsonSMB13}.  

Recently, Holden et al. \cite{journals/corr/abs-2406-19508} uses LLMs to classify Java code defects and achieved good results in the field of software engineering. This work inspired us to think about using LLM as a Linting tool for defect detection of Verilog code and reducing the generation of false positives.


\begin{figure}[t]
	\centering
	\includegraphics[width=0.45\textwidth]{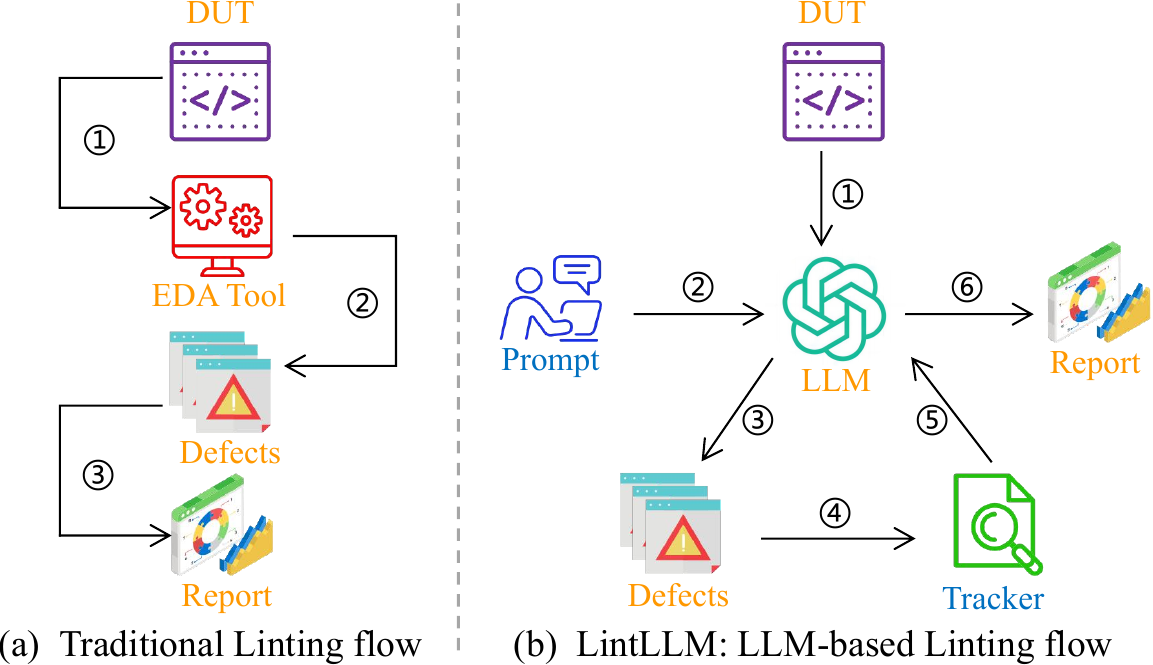} 
	\caption{Workflow of Traditional Linting and LLM-based Linting}
	\label{pic1}
	\vspace{-2mm}
\end{figure}

\section{Methods}

\subsection{Overview}

The traditional Linting flow is shown in Fig. \ref{pic1}(a). To explore the application of LLMs in code defect detection, we proposed the LLM-based Linting flow named LintLLM, as shown in Fig. \ref{pic1}(b). \ding{172}The DUT is input into LLM. \ding{173}The Prompt is used to guide LLM in detecting defects. \ding{174}Many defects are passed to \ding{175}the Tracker for tracking. \ding{176}The main defect is located through the feedback loop. \ding{177}Outputting the detection report. The following methods are used to enhance LLM.



\subsection{Prompt of Logic-Tree \label{method1}}

\begin{figure}[htbp]
	\centering
	\includegraphics[width=0.49\textwidth]{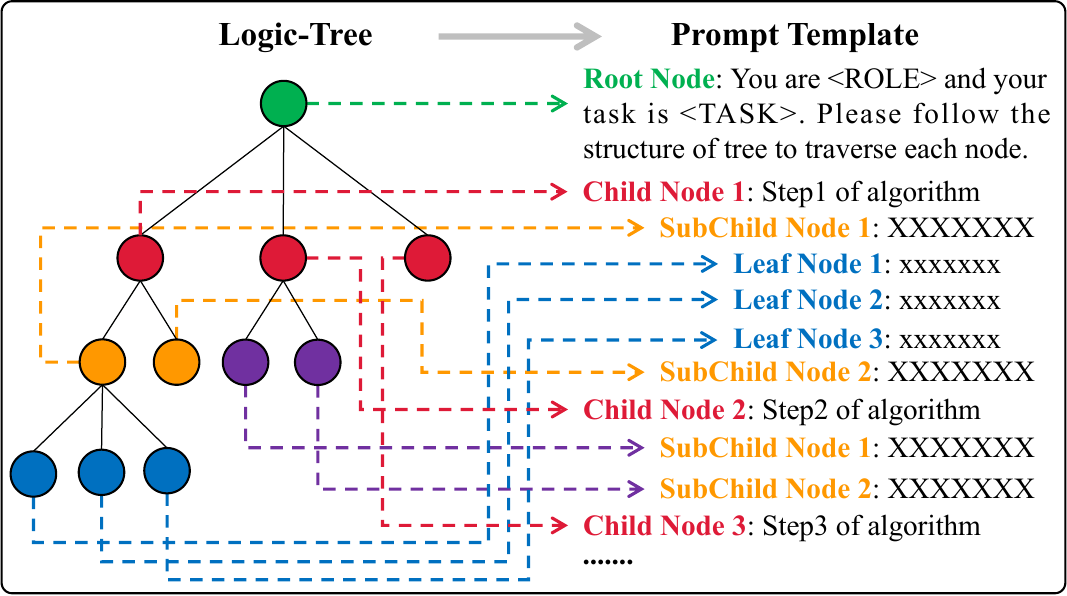} 
	\caption{Template for Prompt of Logic-Tree}
	\label{pic2}
\end{figure}

Clear logic and detailed prompts can be better understood by LLMs. To achieve this, we propose the Prompt of Logic-Tree by drawing ideas from the Tree of Thoughts (ToT) \cite{conf/nips/YaoYZS00N23}, as shown in Fig. \ref{pic2}. Different from ToT, this is a new prompt template that translates complex algorithms for problem-solving into a clear tree structure. This template consists of two parts: the root node specifying the role and task and child nodes describing detailed algorithms for solving the task.

\textbf{Root Node.} The role and task of LLMs are described as the root node. By setting a specific role, LLMs can more accurately answer user questions or perform specific tasks. This role definition helps ensure that the LLM's output aligns with its role positioning, avoiding overly free or off-topic responses. LLMs treat the assigned task as an objective that must be followed throughout the task execution process.

\textbf{Child Nodes.} The characteristic of the Logic-Tree is that the each node of the tree is ordered from left to right and cannot be interchanged. Therefore, it is reasonable to describe the task-solving algorithms with rigorous logic as child nodes. The parent nodes represent the main steps of the solving algorithm, and the child nodes from left to right represent the sequential sub-steps under the main steps.

\subsection{Defect Tracker \label{method2}}

The difference from natural language is that Verilog code is highly logical, and the main defect in the code may cause multiple related secondary defects, as shown in Fig. \ref{pic4}. Due to the dependencies between these defects, Linting tools face challenges in detecting and locating the main defect. Traditional Linting tools can usually only detect line by line and cannot accurately locate the root cause of secondary defects.

\begin{figure}[hb]
\begin{tcolorbox}[colback=white, colframe=white]
\begin{lstlisting}[style=mystyle1,xleftmargin=-0.5em,xrightmargin=-1em,aboveskip=-1em,belowskip=-1.5em]
module complex_1(
    output reg [15:0] qo,
    input [15:0] din,
    input load
);
    reg [7:0] temp_reg; // main defect
	    // [7:0]-->[15:0]
    always @(posedge load) begin
        temp_reg <= din; // secondary defect 1 
        qo <= temp_reg;  // secondary defect 2
    end
endmodule
\end{lstlisting}
\end{tcolorbox}
\caption{The main defect of incorrect bit-width of `temp\_reg' in Line 6 simultaneously causes secondary defects in Line 9 and Line 10.}
\label{pic4}
\end{figure}

To address this scenario, we introduce the Defect Tracker. It guides LLMs to identify the main defect in complex code defects. The specific algorithm is shown in Alg. \ref{defect_tracker}, which mainly includes three steps. 

(\rmnum{1}) Defining the initial defect set. Using LLM to detect all defects in the code, and record the location, type, and dependencies of each defect.

(\rmnum{2}) Gradually fix and re-detect. Defect Tracker will fix each defect one by one, and then re-run LLM to detect the number of remaining defects.

(\rmnum{3}) Judging the minimum number of defects. When the number of remaining defects reaches the minimum value after repairing a certain defect, it indicates that the defect is the main cause of multiple secondary defects.

\begin{algorithm}
	\fontsize{7pt}{9.1pt}\selectfont
	\caption{Defect Tracker Algorithm}
	\label{defect_tracker}
	\begin{algorithmic}[1]
			\REQUIRE A set of detected defects $D = \{D_1, D_2, \ldots, D_m\}$, where $m$ is the number of defects.
			\ENSURE Identify the main defect $D_{{main}}$.
			
			\STATE Initialize a set for the number of remaining defects $R = \{R_1, R_2, \ldots, R_m\}$ after fixing $D$, where $m$ is the number of defects.
			
			\FOR{$i = 1$ to $m$}
			\STATE Fix defect $D_i$ while ignoring the other defects $D_j$ for $j \neq i$.
			\STATE Run the defect detection process, and record the number of remaining defect as $R_i$.
			\ENDFOR
			
			\STATE Find the minimum value in the set $R$, i.e., $R_k$.
			
			\STATE The defect $D_k$ corresponding to $R_k$ is identified as the main defect $D_{main}$.
			
			\RETURN $D_{main}$
		\end{algorithmic}
\end{algorithm}

\section{Benchmark of Static Code Analysis}

Our investigation reveals that previous benchmarks \cite{conf/aspdac/LuLZX24,journals/corr/abs-2405-06840} mainly focus on evaluating the correctness of syntax and function for Verilog code generated by LLMs. Therefore, we create a high-quality benchmark to evaluate LLM's ability for detecting Verilog defects through the following workflows.

\subsection{Correct Code Collection}
Firstly, we screen 150 common Verilog files from Github. Each file is a complete `module-endmodule' block and does not contain the `include' keyword. Then, we delete all comments, aiming to evaluate LLM's understanding and defect detection capabilities only for code. To ensure the correctness of these codes, we verify their syntax and function. Finally, 90 correct design files are obtained.

\subsection{Code Defect Injection \label{method3}}

We define multiple rules through mutation-based defect injection technique \cite{conf/mbmv/Ahmadi-PourHD21}, as shown in Table \ref{rule}. This method imitates the defect that hardware designers unintentionally introduce during the coding process.

\begin{table}[htbp]
	\caption{Rules for code defect injection }
	\vspace{-1mm}
	\centering
	\setlength{\tabcolsep}{2.5pt} 
	\fontsize{7pt}{1pt}\selectfont
	\begin{tabular}{c|>{\raggedright\arraybackslash}p{4.4cm}|>{\raggedright\arraybackslash}p{3.4cm}} 
		\toprule 
		\textbf{Type}\vspace{-0.2pt}   & \multicolumn{1}{c|}{\textbf{Description}}    &  \multicolumn{1}{c}{\textbf{Example}}\\
		\midrule
		1         & replace the reserved keywords           & swap (else if) with (elif)   \\  \rule{0pt}{8pt}
		2         & swap blocking and non-blocking operators    & swap ($=$)  with ($<=$)   \\ \rule{0pt}{8pt}
		3         & swap assignment and relational operators           & swap ($=$) with ($==$)  \\ \rule{0pt}{8pt}
		4         & swap port type of signals                 & swap (input) with (output)  \\ \rule{0pt}{8pt}
		5         & swap data type of signals     & swap (reg) with (wire)  \\  \rule{0pt}{8pt}
		6         & change the bit-width of signals           & swap ([7:0]) with ([15:0])   \\  \rule{0pt}{8pt}
		7         & swap signal edges in the sensitivity list      & swap (posedge) with (negedge) \\ \rule{0pt}{8pt}
		8         & swap logical and bitwise operators        & swap (\&) with (\&\&) \\ \rule{0pt}{8pt}
		9         & change connection symbol in sensitivity list & swap (or) with (\textbar, \textbar\textbar) \\ \rule{0pt}{8pt}
		10        & use undefined or undeclared signal   & undefined signal \\  \rule{0pt}{8pt}
		11        & inject statements that create race or hazard & write-write race for signal out \\ \rule{0pt}{8pt}
		12        & inject statements that cannot be synthesized & unknown or high-impedance state \\ \rule{0pt}{8pt}
		13        & inject defects caused by module instances & signal port floating
		\\ 
		\bottomrule
	\end{tabular}
	\label{rule}
\end{table}

\subsection{Benchmark Construction}

We create an anonymously released benchmark, which contains 11 categories of defects and 90 design files. It includes defect codes and the corresponding line number, as shown in TABLE \ref{benchmark}. Due to the length limit of the paper, the details of the benchmark can be accessed through this link: https://github.com/fangzhigang32/Static-Verilog-Analysis

\begin{table}[htbp]
	\caption{Benchmark for Verilog code defects}
	\centering
	\setlength{\tabcolsep}{16.4pt} 
	\fontsize{7pt}{1pt}\selectfont
	\vspace{-1mm}
	\begin{tabular}{l|l|c} 
		\toprule 
		\textbf{Difficulty Levels}  \vspace{-0.2pt}                                               & \textbf{Defect Categories} \vspace{-0.2pt}            & \multicolumn{1}{c}{\textbf{Count}}  \vspace{-0.2pt}\\ 
		\midrule 
		& Syntax Structure            & 6                       \\ \rule{0pt}{8pt}
		& Signal Usage                & 13                       \\ \rule{0pt}{8pt}
		Simple (30)                                                      & Sensitivity List            & 16                      \\ \rule{0pt}{8pt}
		& Reserved words              & 3                       \\ \rule{0pt}{8pt}
		& Race or Hazard              & 11                       \\ \rule{0pt}{8pt}
		Medium (30)                                                      & Port Type                   & 4                       \\ \rule{0pt}{8pt}
		& Operators                   & 7                        \\ \rule{0pt}{8pt}
		& Module Instances            & 8                        \\ \rule{0pt}{8pt}
		Complex (30)                                                     & Logic Synthesis             & 6                      \\ \rule{0pt}{8pt}
		& Combinational or Sequential & 2                         \\ \rule{0pt}{8pt}
		& Bit width Usage             & 14                        \\
		\midrule  
		\multicolumn{2}{l}{Total}                                                                  & 90               \\ 
		\bottomrule                                         
	\end{tabular}
	\label{benchmark}
\end{table}

\section{Experiment}
\subsection{LLMs Selection} 
We select recent LLMs to facilitate academic research and industrial applications, and TABLE \ref{llm_summary} describes the details of the models. Considering the actual operating environment of code defect detection, the hyper-parameter \textbf{Temperature} is set to 0 to ensure the determinism of LLMs outputs. Note that o1-mini does not support this property.

\begin{table}[htbp]
	\caption{Summary of LLMs Used in Current Study}
	\vspace{-1mm}
	\centering
	\setlength{\tabcolsep}{9.8pt} 
	\fontsize{7pt}{8pt}\selectfont
	\begin{tabular}{l|c|c|cc} 
		\toprule 
		\textbf{Model Name}\vspace{-0.2pt}    & \textbf{Parameters}\vspace{-0.2pt}   & \textbf{Context Length}\vspace{-0.2pt} & \textbf{Release Date}\vspace{-0.2pt} \\
		\midrule \rule{0pt}{8pt}
		GPT-4         & Unintroduced & 8K             & Mar, 2023   \\  \rule{0pt}{8pt}
		Llama-3.1     & 405B         & 128K           & Jul, 2024   \\  \rule{0pt}{8pt}
		GPT-4o        & Unintroduced & 128K           & May, 2024   \\  \rule{0pt}{8pt}
		DeepSeek V2.5 & 236B         & 128K           & Sept, 2024  \\  \rule{0pt}{8pt}
		o1-mini       & Unintroduced & 128K           & Sept, 2024  \\
		\bottomrule
	\end{tabular}
	\label{llm_summary}
\end{table}

The best performing \textbf{commercial EDA} and \textbf{Verilator} \cite{Verilator} are our baselines.

\subsection{Experimental Setting}

To evaluate LLM's ability for detecting Verilog defects, we set up four groups of experiments.

(\rmnum{1}) \textbf{Original}. Directly input DUT into LLM for exploring LLM's basic capability, as shown in Fig. \ref{pic1}(b) (steps \ding{172}$\rightarrow$\ding{177}).

(\rmnum{2}) \textbf{+Prompt of Logic-Tree}. Only Prompt of Logic-Tree is used to explore its impact on LLM in improving the correct rate of defects, as shown in Fig. \ref{pic1}(b) (steps \ding{172}$\rightarrow$\ding{173}$\rightarrow$\ding{177}).

(\rmnum{3}) \textbf{+Defect Tracker}. Only Defect Tracker is used to explore its impact on LLM in reducing the false-positive rate of defects, as shown in Fig. \ref{pic1}(b) (steps \ding{172}$\rightarrow$\ding{174}$\rightarrow$\ding{175}$\rightarrow$\ding{176}$\rightarrow$\ding{177}).

(\rmnum{4}) \textbf{LintLLM}. Prompt of Logic-Tree and Defect Tracker are simultaneously used to explore the LLM's best performance, as shown in Fig. \ref{pic1}(b) (steps \ding{172}$\rightarrow$\ding{173}$\rightarrow$\ding{174}$\rightarrow$\ding{175}$\rightarrow$\ding{176}$\rightarrow$\ding{177}).

\subsection{Evaluation Metrics}
There are two cases for the evaluation of experimental results. (\rmnum{1}) When the defective line in the defect report is consistent with the line number where the defect is injected, it is considered a correct detection.  (\rmnum{2}) When a code line without defect is reported as a defect, it is considered a false positive. To comprehensively evaluate the capability of code defect detection by LLMs, we defined two metrics. \textbf{Correct Rate (CR)}: the percentage of defects correctly detected in all DUTs. \textbf{False-Positive Rate (FR)}: the percentage of false positive defects in all DUTs. The CR is better when higher (\textcolor{blue}{$\nearrow$}), and FR is better when lower (\textcolor{red}{$\searrow$}).

\subsection{Research Questions} 
This paper evaluates the effectiveness of LLMs in code defect detection and examines the impact of our proposed methods. It includes an investigation structured around three specific Research Questions (RQs).

\begin{itemize}
\item
\textbf{RQ1:} What are the differences in detection performance between LLMs and EDA tools?
\item
\textbf{RQ2:} Why our methods can improve the detection performance of LLMs?
\item
\textbf{RQ3:} What are the advantages of LLMs in detection costs compared with commercial EDA tools?
\end{itemize}

\section{Result}


The experiment was conducted on the proposed benchmark, with the complete results presented in TABLE \ref{main_result}. The results demonstrate that the correct rate of LLMs significantly outperforms EDA tools, especially when enhanced with Prompt of Logic-Tree and Defect Tracker. Among the evaluated LLMs, o1-mini achieves the highest correct rate of 83.33\% and the lowest false-positive rate of 12.22\%. Compared to the best performing commercial EDA tool, o1-mini's correct rate increased by 18.89\% and false-positive rate decreased by 15.56\%. When compared to Verilator, its correct rate increased by 21.11\% and false-positive rate decreased by 20.00\%.

\begin{table}[htbp]
	\setlength{\tabcolsep}{0.8pt} 
	\caption{Main Results - The performance of code defect detection tasks using EDA tools and LLMs. \ding{172} is Original. \ding{173} is + Prompt of Logic-Tree. \ding{174} is + Defect Tracker. \ding{175} is LintLLM. o1-mini +LintLLM achieves the best performance, with an correct rate of 83.33\% and a false-positive rate of 12.22\%. }
	\label{main_result}
	\fontsize{7pt}{8pt}\selectfont
	\vspace{-1mm}
	\begin{tabular}{l|cc|cc|cc|cc}
		\toprule
		\multirow{2}{*}{\textbf{ Tools}} &
		\multicolumn{2}{c|}{\ding{172}} &
		\multicolumn{2}{c|}{\ding{173}} &
		\multicolumn{2}{c|}{\ding{174}} &
		\multicolumn{2}{c}{\ding{175}}\\ \cline{2-3} \cline{4-5} \cline{6-7} \cline{8-9} \rule{0pt}{8pt} 
		& \textbf{CR} & \textbf{FR} & \textbf{CR} & \textbf{FR} & \textbf{CR} & \textbf{FR} & \textbf{CR} & \textbf{FR} \\ 
		\midrule \rule{0pt}{8pt}
		Commercial EDA          & \cellcolor{green!27}64.44\%      & \cellcolor{red!35}27.78\%             & \cellcolor{green!15}--            & \cellcolor{red!15}--                   & \cellcolor{green!15}--            & \cellcolor{red!15}--                   & \cellcolor{green!15}--               & \cellcolor{red!15}--                     \\ \rule{0pt}{8pt}
		Verilator                & \cellcolor{green!21}62.22\%      & \cellcolor{red!30}32.22\%    & \cellcolor{green!15}--            & \cellcolor{red!15}--                   & \cellcolor{green!15}--            & \cellcolor{red!15}--                   & \cellcolor{green!15}--               & \cellcolor{red!15}--                     \\ \rule{0pt}{8pt}
		Llama-3.1          & \cellcolor{green!24}63.33\%      & \cellcolor{red!15}61.11\%             & \cellcolor{green!33}66.67\%      & \cellcolor{red!17.5}60.00\%             & \cellcolor{green!33}66.67\%      & \cellcolor{red!30}32.22\%             & \cellcolor{green!39}68.89\%         & \cellcolor{red!32.5}31.11\%               \\ \rule{0pt}{8pt}
		DeepSeek V2.5      & \cellcolor{green!24}63.33\%      & \cellcolor{red!50}20.00\%             & \cellcolor{green!48}\textbf{75.56\%}      & \cellcolor{red!22.5}47.78\%             & \cellcolor{green!51}78.89\%      & \cellcolor{red!47.5}21.11\%             & \cellcolor{green!57}81.11\%         & \cellcolor{red!52.5}18.89\%               \\ \rule{0pt}{8pt}
		GPT-4              & \cellcolor{green!15}47.78\%      & \cellcolor{red!25}36.67\%             & \cellcolor{green!18}48.89\%      & \cellcolor{red!42.5}24.44\%             & \cellcolor{green!21}62.22\%      & \cellcolor{red!52.5}18.89\%             & \cellcolor{green!33}66.67\%         & \cellcolor{red!27.5}33.33\%               \\ \rule{0pt}{8pt}
		GPT-4o             & \cellcolor{green!30}65.56\%      & \cellcolor{red!45}22.22\%             & \cellcolor{green!30}65.56\%      & \cellcolor{red!50}\textbf{20.00\%}    & \cellcolor{green!36}67.78\%      & \cellcolor{red!40}25.56\%             & \cellcolor{green!45}73.33\%         & \cellcolor{red!37.5}26.67\%               \\ 	\rule{0pt}{8pt}		
		o1-mini            & \cellcolor{green!36}\textbf{67.78\%}      & \cellcolor{red!60}\textbf{11.11\%}             & \cellcolor{green!42}70.00\%      & \cellcolor{red!20}55.56\%             & \cellcolor{green!54}\textbf{80.00\%}      & \cellcolor{red!55}\textbf{13.33\%}             & \cellcolor{green!60}\textbf{83.33\%}         & \cellcolor{red!57.5}\textbf{12.22\%}               \\
		\bottomrule
	\end{tabular}			
	\footnotemark \textit{Note:} The highest contrast colors represent the best rate.
\end{table}

Overall, o1-mini shows the best performance, combining a high correct rate with a low false-positive rate, demonstrating the effectiveness of LLMs in code defect detection tasks.

\begin{table*}[htbp]
	\caption{Detailed Results - Defect detection reports using EDA tools and LLMs on the benchmark. \ding{182} is Commercial EDA. \ding{183} is Verilator. \ding{184} is Llama-3.1 +LintLLM. \ding{185} is DeepSeek V2.5 +LintLLM. \ding{186} is GPT-4 +LintLLM. \ding{187} is GPT-4o +LintLLM. \ding{188} is o1-mini +LintLLM. \textbf{C} indicates whether the defect is detected correctly, and \textbf{F} indicates the number of false positives (0 as `-'). It shows that LLMs perform better than EDA tools.}  
	\label{Detail_Results}
	\centering
	\setlength{\tabcolsep}{2.42pt} 
	\fontsize{7pt}{1pt}\selectfont
	\vspace{-1mm}
	\begin{tabular}{c|cccccccccccccc|c|cccccccccccccc|c|cccccccccccccc} 
		\toprule
		\multirow{2}{*}{\textbf{Simple}} &
		\multicolumn{2}{c}{\ding{182}} & \multicolumn{2}{c}{\ding{183}} & \multicolumn{2}{c}{\ding{184}} & \multicolumn{2}{c}{\ding{185}} & \multicolumn{2}{c}{\ding{186}} & \multicolumn{2}{c}{\ding{187}} & \multicolumn{2}{c|}{\ding{188}} & 
		\multirow{2}{*}{\textbf{Medium}} &
		\multicolumn{2}{c}{\ding{182}} & \multicolumn{2}{c}{\ding{183}} & \multicolumn{2}{c}{\ding{184}} & \multicolumn{2}{c}{\ding{185}} & \multicolumn{2}{c}{\ding{186}} & \multicolumn{2}{c}{\ding{187}} & \multicolumn{2}{c|}{\ding{188}}  & 
		\multirow{2}{*}{\textbf{Complex}} &
		\multicolumn{2}{c}{\ding{182}} & \multicolumn{2}{c}{\ding{183}} & \multicolumn{2}{c}{\ding{184}} & \multicolumn{2}{c}{\ding{185}} & \multicolumn{2}{c}{\ding{186}} & \multicolumn{2}{c}{\ding{187}} & \multicolumn{2}{c}{\ding{188}}  \\
		\cline{2-15} \cline{17-30} \cline{32-45}  \rule{0pt}{8pt}
		& \textbf{C} & \textbf{F} & \textbf{C} & \textbf{F} & \textbf{C} & \textbf{F} & \textbf{C} & \textbf{F} & \textbf{C} & \textbf{F} & \textbf{C} & \textbf{F} & \textbf{C} & \textbf{F} & &  \textbf{C} & \textbf{F} & \textbf{C} & \textbf{F} & \textbf{C} & \textbf{F} & \textbf{C} & \textbf{F} & \textbf{C} & \textbf{F} & \textbf{C} & \textbf{F} & \textbf{C} & \textbf{F} &  & \textbf{C} & \textbf{F} & \textbf{C} & \textbf{F} & \textbf{C} & \textbf{F} & \textbf{C} & \textbf{F} & \textbf{C} & \textbf{F} & \textbf{C} & \textbf{F} & \textbf{C} & \textbf{F}\\
		
		\midrule \rule{0pt}{8pt}
		s01	&	\cellcolor{green!45}{\checked}	&	-	&	\cellcolor{green!45}{\checked}	&	-	&	\cellcolor{green!45}{\checked}	&	-	&	\cellcolor{green!45}{\checked}	&	-	&	\cellcolor{green!45}{\checked}	&	-	&	\cellcolor{green!45}{\checked}	&	-	&	\cellcolor{green!45}{\checked}	&	-	&	m01	&	\cellcolor{green!45}{\checked}	&	1	&	\cellcolor{green!45}{\checked}	&	1	&	\cellcolor{green!45}{\checked}	&	-	&	\cellcolor{green!45}{\checked}	&	-	&	\cellcolor{green!45}{\checked}	&	-	&	\cellcolor{green!45}{\checked}	&	-	&	\cellcolor{green!45}{\checked}	&	-	&	c01	&	\cellcolor{red!45}{$\times$}	&	-	&	\cellcolor{green!45}{\checked}	&	1	&	\cellcolor{green!45}{\checked}	&	-	&	\cellcolor{red!45}{$\times$}	&	1	&	\cellcolor{green!45}{\checked}	&	-	&	\cellcolor{green!45}{\checked}	&	-	&	\cellcolor{green!45}{\checked}	&	-
		\\ \rule{0pt}{8pt} 																																																																																								
		s02	&	\cellcolor{green!45}{\checked}	&	-	&	\cellcolor{green!45}{\checked}	&	-	&	\cellcolor{green!45}{\checked}	&	-	&	\cellcolor{green!45}{\checked}	&	-	&	\cellcolor{green!45}{\checked}	&	-	&	\cellcolor{green!45}{\checked}	&	-	&	\cellcolor{green!45}{\checked}	&	-	&	m02	&	\cellcolor{red!45}{$\times$}	&	-	&	\cellcolor{green!45}{\checked}	&	-	&	\cellcolor{green!45}{\checked}	&	-	&	\cellcolor{green!45}{\checked}	&	-	&	\cellcolor{green!45}{\checked}	&	-	&	\cellcolor{green!45}{\checked}	&	-	&	\cellcolor{green!45}{\checked}	&	-	&	c02	&	\cellcolor{red!45}{$\times$}	&	3	&	\cellcolor{green!45}{\checked}	&	-	&	\cellcolor{green!45}{\checked}	&	-	&	\cellcolor{green!45}{\checked}	&	-	&	\cellcolor{red!45}{$\times$}	&	1	&	\cellcolor{red!45}{$\times$}	&	1	&	\cellcolor{green!45}{\checked}	&	-
		\\ \rule{0pt}{8pt} 																																																																																								
		s03	&	\cellcolor{green!45}{\checked}	&	-	&	\cellcolor{green!45}{\checked}	&	-	&	\cellcolor{green!45}{\checked}	&	-	&	\cellcolor{green!45}{\checked}	&	-	&	\cellcolor{green!45}{\checked}	&	-	&	\cellcolor{green!45}{\checked}	&	-	&	\cellcolor{green!45}{\checked}	&	-	&	m03	&	\cellcolor{green!45}{\checked}	&	-	&	\cellcolor{green!45}{\checked}	&	2	&	\cellcolor{green!45}{\checked}	&	-	&	\cellcolor{green!45}{\checked}	&	-	&	\cellcolor{green!45}{\checked}	&	-	&	\cellcolor{green!45}{\checked}	&	-	&	\cellcolor{green!45}{\checked}	&	-	&	c03	&	\cellcolor{green!45}{\checked}	&	-	&	\cellcolor{green!45}{\checked}	&	-	&	\cellcolor{red!45}{$\times$}	&	1	&	\cellcolor{red!45}{$\times$}	&	1	&	\cellcolor{green!45}{\checked}	&	-	&	\cellcolor{green!45}{\checked}	&	-	&	\cellcolor{green!45}{\checked}	&	-
		\\ \rule{0pt}{8pt} 																																																																																								
		s04	&	\cellcolor{green!45}{\checked}	&	-	&	\cellcolor{green!45}{\checked}	&	-	&	\cellcolor{green!45}{\checked}	&	-	&	\cellcolor{green!45}{\checked}	&	-	&	\cellcolor{green!45}{\checked}	&	-	&	\cellcolor{green!45}{\checked}	&	-	&	\cellcolor{green!45}{\checked}	&	-	&	m04	&	\cellcolor{green!45}{\checked}	&	-	&	\cellcolor{red!45}{$\times$}	&	-	&	\cellcolor{red!45}{$\times$}	&	1	&	\cellcolor{green!45}{\checked}	&	-	&	\cellcolor{red!45}{$\times$}	&	1	&	\cellcolor{green!45}{\checked}	&	-	&	\cellcolor{green!45}{\checked}	&	-	&	c04	&	\cellcolor{red!45}{$\times$}	&	-	&	\cellcolor{red!45}{$\times$}	&	-	&	\cellcolor{green!45}{\checked}	&	-	&	\cellcolor{red!45}{$\times$}	&	1	&	\cellcolor{green!45}{\checked}	&	-	&	\cellcolor{green!45}{\checked}	&	-	&	\cellcolor{green!45}{\checked}	&	-
		\\ \rule{0pt}{8pt} 																																																																																								
		s05	&	\cellcolor{green!45}{\checked}	&	-	&	\cellcolor{green!45}{\checked}	&	1	&	\cellcolor{green!45}{\checked}	&	-	&	\cellcolor{green!45}{\checked}	&	-	&	\cellcolor{green!45}{\checked}	&	-	&	\cellcolor{green!45}{\checked}	&	-	&	\cellcolor{green!45}{\checked}	&	-	&	m05	&	\cellcolor{green!45}{\checked}	&	-	&	\cellcolor{green!45}{\checked}	&	-	&	\cellcolor{red!45}{$\times$}	&	1	&	\cellcolor{green!45}{\checked}	&	-	&	\cellcolor{green!45}{\checked}	&	-	&	\cellcolor{green!45}{\checked}	&	-	&	\cellcolor{green!45}{\checked}	&	-	&	c05	&	\cellcolor{green!45}{\checked}	&	-	&	\cellcolor{green!45}{\checked}	&	2	&	\cellcolor{red!45}{$\times$}	&	1	&	\cellcolor{red!45}{$\times$}	&	1	&	\cellcolor{red!45}{$\times$}	&	1	&	\cellcolor{red!45}{$\times$}	&	1	&	\cellcolor{red!45}{$\times$}	&	1
		\\ \rule{0pt}{8pt} 																																																																																								
		s06	&	\cellcolor{green!45}{\checked}	&	-	&	\cellcolor{green!45}{\checked}	&	-	&	\cellcolor{green!45}{\checked}	&	-	&	\cellcolor{green!45}{\checked}	&	-	&	\cellcolor{green!45}{\checked}	&	-	&	\cellcolor{green!45}{\checked}	&	-	&	\cellcolor{green!45}{\checked}	&	-	&	m06	&	\cellcolor{red!45}{$\times$}	&	-	&	\cellcolor{green!45}{\checked}	&	-	&	\cellcolor{red!45}{$\times$}	&	1	&	\cellcolor{green!45}{\checked}	&	-	&	\cellcolor{green!45}{\checked}	&	-	&	\cellcolor{green!45}{\checked}	&	-	&	\cellcolor{green!45}{\checked}	&	-	&	c06	&	\cellcolor{green!45}{\checked}	&	-	&	\cellcolor{green!45}{\checked}	&	-	&	\cellcolor{red!45}{$\times$}	&	1	&	\cellcolor{red!45}{$\times$}	&	1	&	\cellcolor{green!45}{\checked}	&	-	&	\cellcolor{green!45}{\checked}	&	-	&	\cellcolor{green!45}{\checked}	&	-
		\\ \rule{0pt}{8pt} 																																																																																								
		s07	&	\cellcolor{green!45}{\checked}	&	-	&	\cellcolor{green!45}{\checked}	&	-	&	\cellcolor{green!45}{\checked}	&	-	&	\cellcolor{green!45}{\checked}	&	-	&	\cellcolor{green!45}{\checked}	&	-	&	\cellcolor{green!45}{\checked}	&	-	&	\cellcolor{green!45}{\checked}	&	-	&	m07	&	\cellcolor{red!45}{$\times$}	&	-	&	\cellcolor{green!45}{\checked}	&	-	&	\cellcolor{red!45}{$\times$}	&	1	&	\cellcolor{green!45}{\checked}	&	-	&	\cellcolor{green!45}{\checked}	&	-	&	\cellcolor{green!45}{\checked}	&	-	&	\cellcolor{green!45}{\checked}	&	-	&	c07	&	\cellcolor{red!45}{$\times$}	&	-	&	\cellcolor{red!45}{$\times$}	&	-	&	\cellcolor{red!45}{$\times$}	&	1	&	\cellcolor{green!45}{\checked}	&	-	&	\cellcolor{green!45}{\checked}	&	-	&	\cellcolor{green!45}{\checked}	&	-	&	\cellcolor{green!45}{\checked}	&	-
		\\ \rule{0pt}{8pt} 																																																																																								
		s08	&	\cellcolor{green!45}{\checked}	&	-	&	\cellcolor{green!45}{\checked}	&	-	&	\cellcolor{green!45}{\checked}	&	-	&	\cellcolor{green!45}{\checked}	&	-	&	\cellcolor{green!45}{\checked}	&	-	&	\cellcolor{green!45}{\checked}	&	-	&	\cellcolor{green!45}{\checked}	&	-	&	m08	&	\cellcolor{red!45}{$\times$}	&	-	&	\cellcolor{red!45}{$\times$}	&	1	&	\cellcolor{red!45}{$\times$}	&	1	&	\cellcolor{green!45}{\checked}	&	-	&	\cellcolor{green!45}{\checked}	&	-	&	\cellcolor{green!45}{\checked}	&	-	&	\cellcolor{green!45}{\checked}	&	-	&	c08	&	\cellcolor{green!45}{\checked}	&	-	&	\cellcolor{green!45}{\checked}	&	-	&	\cellcolor{red!45}{$\times$}	&	1	&	\cellcolor{red!45}{$\times$}	&	1	&	\cellcolor{red!45}{$\times$}	&	1	&	\cellcolor{red!45}{$\times$}	&	1	&	\cellcolor{green!45}{\checked}	&	-
		\\ \rule{0pt}{8pt} 																																																																																								
		s09	&	\cellcolor{green!45}{\checked}	&	-	&	\cellcolor{green!45}{\checked}	&	-	&	\cellcolor{green!45}{\checked}	&	-	&	\cellcolor{green!45}{\checked}	&	-	&	\cellcolor{green!45}{\checked}	&	-	&	\cellcolor{green!45}{\checked}	&	-	&	\cellcolor{green!45}{\checked}	&	-	&	m09	&	\cellcolor{green!45}{\checked}	&	-	&	\cellcolor{green!45}{\checked}	&	1	&	\cellcolor{red!45}{$\times$}	&	1	&	\cellcolor{green!45}{\checked}	&	-	&	\cellcolor{green!45}{\checked}	&	-	&	\cellcolor{green!45}{\checked}	&	-	&	\cellcolor{green!45}{\checked}	&	-	&	c09	&	\cellcolor{red!45}{$\times$}	&	1	&	\cellcolor{red!45}{$\times$}	&	-	&	\cellcolor{red!45}{$\times$}	&	1	&	\cellcolor{red!45}{$\times$}	&	1	&	\cellcolor{red!45}{$\times$}	&	1	&	\cellcolor{red!45}{$\times$}	&	1	&	\cellcolor{red!45}{$\times$}	&	1
		\\ \rule{0pt}{8pt} 																																																																																								
		s10	&	\cellcolor{green!45}{\checked}	&	-	&	\cellcolor{green!45}{\checked}	&	-	&	\cellcolor{green!45}{\checked}	&	-	&	\cellcolor{green!45}{\checked}	&	-	&	\cellcolor{green!45}{\checked}	&	-	&	\cellcolor{green!45}{\checked}	&	-	&	\cellcolor{green!45}{\checked}	&	-	&	m10	&	\cellcolor{green!45}{\checked}	&	-	&	\cellcolor{green!45}{\checked}	&	1	&	\cellcolor{red!45}{$\times$}	&	1	&	\cellcolor{green!45}{\checked}	&	-	&	\cellcolor{green!45}{\checked}	&	-	&	\cellcolor{green!45}{\checked}	&	-	&	\cellcolor{green!45}{\checked}	&	-	&	c10	&	\cellcolor{red!45}{$\times$}	&	-	&	\cellcolor{red!45}{$\times$}	&	1	&	\cellcolor{green!45}{\checked}	&	-	&	\cellcolor{green!45}{\checked}	&	-	&	\cellcolor{green!45}{\checked}	&	-	&	\cellcolor{red!45}{$\times$}	&	1	&	\cellcolor{red!45}{$\times$}	&	1
		\\ \rule{0pt}{8pt} 																																																																																								
		s11	&	\cellcolor{green!45}{\checked}	&	-	&	\cellcolor{green!45}{\checked}	&	-	&	\cellcolor{green!45}{\checked}	&	-	&	\cellcolor{red!45}{$\times$}	&	1	&	\cellcolor{red!45}{$\times$}	&	1	&	\cellcolor{green!45}{\checked}	&	-	&	\cellcolor{green!45}{\checked}	&	-	&	m11	&	\cellcolor{green!45}{\checked}	&	-	&	\cellcolor{red!45}{$\times$}	&	-	&	\cellcolor{green!45}{\checked}	&	-	&	\cellcolor{green!45}{\checked}	&	-	&	\cellcolor{red!45}{$\times$}	&	1	&	\cellcolor{red!45}{$\times$}	&	1	&	\cellcolor{green!45}{\checked}	&	-	&	c11	&	\cellcolor{red!45}{$\times$}	&	-	&	\cellcolor{red!45}{$\times$}	&	-	&	\cellcolor{green!45}{\checked}	&	-	&	\cellcolor{green!45}{\checked}	&	-	&	\cellcolor{green!45}{\checked}	&	-	&	\cellcolor{green!45}{\checked}	&	-	&	\cellcolor{green!45}{\checked}	&	-
		\\ \rule{0pt}{8pt} 																																																																																								
		s12	&	\cellcolor{green!45}{\checked}	&	-	&	\cellcolor{green!45}{\checked}	&	-	&	\cellcolor{green!45}{\checked}	&	-	&	\cellcolor{green!45}{\checked}	&	-	&	\cellcolor{green!45}{\checked}	&	-	&	\cellcolor{green!45}{\checked}	&	-	&	\cellcolor{green!45}{\checked}	&	-	&	m12	&	\cellcolor{red!45}{$\times$}	&	-	&	\cellcolor{green!45}{\checked}	&	-	&	\cellcolor{red!45}{$\times$}	&	1	&	\cellcolor{green!45}{\checked}	&	-	&	\cellcolor{green!45}{\checked}	&	-	&	\cellcolor{green!45}{\checked}	&	-	&	\cellcolor{green!45}{\checked}	&	-	&	c12	&	\cellcolor{red!45}{$\times$}	&	-	&	\cellcolor{red!45}{$\times$}	&	-	&	\cellcolor{green!45}{\checked}	&	-	&	\cellcolor{green!45}{\checked}	&	-	&	\cellcolor{green!45}{\checked}	&	-	&	\cellcolor{red!45}{$\times$}	&	1	&	\cellcolor{green!45}{\checked}	&	-
		\\ \rule{0pt}{8pt} 																																																																																								
		s13	&	\cellcolor{green!45}{\checked}	&	-	&	\cellcolor{green!45}{\checked}	&	-	&	\cellcolor{green!45}{\checked}	&	-	&	\cellcolor{green!45}{\checked}	&	-	&	\cellcolor{green!45}{\checked}	&	-	&	\cellcolor{green!45}{\checked}	&	-	&	\cellcolor{green!45}{\checked}	&	-	&	m13	&	\cellcolor{green!45}{\checked}	&	-	&	\cellcolor{green!45}{\checked}	&	-	&	\cellcolor{green!45}{\checked}	&	-	&	\cellcolor{green!45}{\checked}	&	-	&	\cellcolor{green!45}{\checked}	&	-	&	\cellcolor{green!45}{\checked}	&	-	&	\cellcolor{green!45}{\checked}	&	-	&	c13	&	\cellcolor{red!45}{$\times$}	&	-	&	\cellcolor{red!45}{$\times$}	&	-	&	\cellcolor{red!45}{$\times$}	&	1	&	\cellcolor{red!45}{$\times$}	&	1	&	\cellcolor{red!45}{$\times$}	&	-	&	\cellcolor{red!45}{$\times$}	&	1	&	\cellcolor{red!45}{$\times$}	&	1
		\\ \rule{0pt}{8pt} 																																																																																								
		s14	&	\cellcolor{green!45}{\checked}	&	-	&	\cellcolor{green!45}{\checked}	&	-	&	\cellcolor{green!45}{\checked}	&	-	&	\cellcolor{green!45}{\checked}	&	-	&	\cellcolor{green!45}{\checked}	&	-	&	\cellcolor{green!45}{\checked}	&	-	&	\cellcolor{green!45}{\checked}	&	-	&	m14	&	\cellcolor{red!45}{$\times$}	&	-	&	\cellcolor{green!45}{\checked}	&	-	&	\cellcolor{green!45}{\checked}	&	-	&	\cellcolor{green!45}{\checked}	&	-	&	\cellcolor{red!45}{$\times$}	&	1	&	\cellcolor{red!45}{$\times$}	&	1	&	\cellcolor{green!45}{\checked}	&	-	&	c14	&	\cellcolor{green!45}{\checked}	&	-	&	\cellcolor{green!45}{\checked}	&	2	&	\cellcolor{red!45}{$\times$}	&	1	&	\cellcolor{green!45}{\checked}	&	-	&	\cellcolor{red!45}{$\times$}	&	1	&	\cellcolor{red!45}{$\times$}	&	1	&	\cellcolor{green!45}{\checked}	&	-
		\\ \rule{0pt}{8pt} 																																																																																								
		s15	&	\cellcolor{green!45}{\checked}	&	-	&	\cellcolor{green!45}{\checked}	&	-	&	\cellcolor{green!45}{\checked}	&	-	&	\cellcolor{green!45}{\checked}	&	-	&	\cellcolor{green!45}{\checked}	&	-	&	\cellcolor{green!45}{\checked}	&	-	&	\cellcolor{green!45}{\checked}	&	-	&	m15	&	\cellcolor{green!45}{\checked}	&	-	&	\cellcolor{red!45}{$\times$}	&	-	&	\cellcolor{green!45}{\checked}	&	-	&	\cellcolor{green!45}{\checked}	&	-	&	\cellcolor{green!45}{\checked}	&	-	&	\cellcolor{green!45}{\checked}	&	-	&	\cellcolor{green!45}{\checked}	&	-	&	c15	&	\cellcolor{red!45}{$\times$}	&	-	&	\cellcolor{red!45}{$\times$}	&	-	&	\cellcolor{red!45}{$\times$}	&	1	&	\cellcolor{red!45}{$\times$}	&	1	&	\cellcolor{red!45}{$\times$}	&	1	&	\cellcolor{green!45}{\checked}	&	-	&	\cellcolor{red!45}{$\times$}	&	1
		\\ \rule{0pt}{8pt} 																																																																																								
		s16	&	\cellcolor{green!45}{\checked}	&	-	&	\cellcolor{red!45}{$\times$}	&	-	&	\cellcolor{green!45}{\checked}	&	-	&	\cellcolor{green!45}{\checked}	&	-	&	\cellcolor{green!45}{\checked}	&	-	&	\cellcolor{green!45}{\checked}	&	-	&	\cellcolor{green!45}{\checked}	&	-	&	m16	&	\cellcolor{red!45}{$\times$}	&	3	&	\cellcolor{green!45}{\checked}	&	-	&	\cellcolor{green!45}{\checked}	&	-	&	\cellcolor{green!45}{\checked}	&	-	&	\cellcolor{green!45}{\checked}	&	-	&	\cellcolor{green!45}{\checked}	&	-	&	\cellcolor{green!45}{\checked}	&	-	&	c16	&	\cellcolor{red!45}{$\times$}	&	-	&	\cellcolor{red!45}{$\times$}	&	-	&	\cellcolor{red!45}{$\times$}	&	1	&	\cellcolor{green!45}{\checked}	&	-	&	\cellcolor{red!45}{$\times$}	&	1	&	\cellcolor{red!45}{$\times$}	&	1	&	\cellcolor{green!45}{\checked}	&	-
		\\ \rule{0pt}{8pt} 																																																																																								
		s17	&	\cellcolor{green!45}{\checked}	&	-	&	\cellcolor{red!45}{$\times$}	&	-	&	\cellcolor{green!45}{\checked}	&	-	&	\cellcolor{green!45}{\checked}	&	-	&	\cellcolor{green!45}{\checked}	&	-	&	\cellcolor{green!45}{\checked}	&	-	&	\cellcolor{green!45}{\checked}	&	-	&	m17	&	\cellcolor{red!45}{$\times$}	&	-	&	\cellcolor{red!45}{$\times$}	&	-	&	\cellcolor{green!45}{\checked}	&	-	&	\cellcolor{green!45}{\checked}	&	-	&	\cellcolor{green!45}{\checked}	&	-	&	\cellcolor{green!45}{\checked}	&	-	&	\cellcolor{green!45}{\checked}	&	-	&	c17	&	\cellcolor{red!45}{$\times$}	&	1	&	\cellcolor{red!45}{$\times$}	&	-	&	\cellcolor{red!45}{$\times$}	&	1	&	\cellcolor{red!45}{$\times$}	&	1	&	\cellcolor{red!45}{$\times$}	&	1	&	\cellcolor{red!45}{$\times$}	&	1	&	\cellcolor{green!45}{\checked}	&	-
		\\ \rule{0pt}{8pt} 																																																																																								
		s18	&	\cellcolor{green!45}{\checked}	&	-	&	\cellcolor{green!45}{\checked}	&	-	&	\cellcolor{green!45}{\checked}	&	-	&	\cellcolor{green!45}{\checked}	&	-	&	\cellcolor{green!45}{\checked}	&	-	&	\cellcolor{green!45}{\checked}	&	-	&	\cellcolor{green!45}{\checked}	&	-	&	m18	&	\cellcolor{red!45}{$\times$}	&	-	&	\cellcolor{red!45}{$\times$}	&	-	&	\cellcolor{green!45}{\checked}	&	-	&	\cellcolor{green!45}{\checked}	&	-	&	\cellcolor{red!45}{$\times$}	&	1	&	\cellcolor{green!45}{\checked}	&	-	&	\cellcolor{green!45}{\checked}	&	-	&	c18	&	\cellcolor{red!45}{$\times$}	&	-	&	\cellcolor{red!45}{$\times$}	&	-	&	\cellcolor{green!45}{\checked}	&	-	&	\cellcolor{green!45}{\checked}	&	-	&	\cellcolor{green!45}{\checked}	&	-	&	\cellcolor{red!45}{$\times$}	&	1	&	\cellcolor{red!45}{$\times$}	&	-
		\\ \rule{0pt}{8pt} 																																																																																								
		s19	&	\cellcolor{green!45}{\checked}	&	-	&	\cellcolor{green!45}{\checked}	&	1	&	\cellcolor{green!45}{\checked}	&	-	&	\cellcolor{green!45}{\checked}	&	-	&	\cellcolor{green!45}{\checked}	&	-	&	\cellcolor{green!45}{\checked}	&	-	&	\cellcolor{green!45}{\checked}	&	-	&	m19	&	\cellcolor{green!45}{\checked}	&	-	&	\cellcolor{red!45}{$\times$}	&	-	&	\cellcolor{green!45}{\checked}	&	-	&	\cellcolor{green!45}{\checked}	&	-	&	\cellcolor{green!45}{\checked}	&	-	&	\cellcolor{green!45}{\checked}	&	-	&	\cellcolor{green!45}{\checked}	&	-	&	c19	&	\cellcolor{green!45}{\checked}	&	-	&	\cellcolor{red!45}{$\times$}	&	-	&	\cellcolor{green!45}{\checked}	&	-	&	\cellcolor{green!45}{\checked}	&	-	&	\cellcolor{red!45}{$\times$}	&	1	&	\cellcolor{red!45}{$\times$}	&	1	&	\cellcolor{green!45}{\checked}	&	-
		\\ \rule{0pt}{8pt} 																																																																																								
		s20	&	\cellcolor{green!45}{\checked}	&	1	&	\cellcolor{green!45}{\checked}	&	-	&	\cellcolor{green!45}{\checked}	&	-	&	\cellcolor{green!45}{\checked}	&	-	&	\cellcolor{green!45}{\checked}	&	-	&	\cellcolor{green!45}{\checked}	&	-	&	\cellcolor{green!45}{\checked}	&	-	&	m20	&	\cellcolor{red!45}{$\times$}	&	2	&	\cellcolor{green!45}{\checked}	&	-	&	\cellcolor{green!45}{\checked}	&	-	&	\cellcolor{green!45}{\checked}	&	-	&	\cellcolor{red!45}{$\times$}	&	2	&	\cellcolor{green!45}{\checked}	&	-	&	\cellcolor{green!45}{\checked}	&	-	&	c20	&	\cellcolor{green!45}{\checked}	&	-	&	\cellcolor{green!45}{\checked}	&	-	&	\cellcolor{green!45}{\checked}	&	-	&	\cellcolor{green!45}{\checked}	&	-	&	\cellcolor{red!45}{$\times$}	&	1	&	\cellcolor{red!45}{$\times$}	&	1	&	\cellcolor{green!45}{\checked}	&	-
		\\ \rule{0pt}{8pt} 																																																																																								
		s21	&	\cellcolor{green!45}{\checked}	&	1	&	\cellcolor{green!45}{\checked}	&	-	&	\cellcolor{green!45}{\checked}	&	-	&	\cellcolor{green!45}{\checked}	&	-	&	\cellcolor{green!45}{\checked}	&	-	&	\cellcolor{green!45}{\checked}	&	-	&	\cellcolor{green!45}{\checked}	&	-	&	m21	&	\cellcolor{green!45}{\checked}	&	-	&	\cellcolor{red!45}{$\times$}	&	3	&	\cellcolor{red!45}{$\times$}	&	1	&	\cellcolor{green!45}{\checked}	&	-	&	\cellcolor{red!45}{$\times$}	&	1	&	\cellcolor{red!45}{$\times$}	&	1	&	\cellcolor{green!45}{\checked}	&	-	&	c21	&	\cellcolor{green!45}{\checked}	&	1	&	\cellcolor{green!45}{\checked}	&	-	&	\cellcolor{red!45}{$\times$}	&	1	&	\cellcolor{red!45}{$\times$}	&	1	&	\cellcolor{red!45}{$\times$}	&	1	&	\cellcolor{red!45}{$\times$}	&	1	&	\cellcolor{red!45}{$\times$}	&	1
		\\ \rule{0pt}{8pt} 																																																																																								
		s22	&	\cellcolor{green!45}{\checked}	&	-	&	\cellcolor{green!45}{\checked}	&	-	&	\cellcolor{green!45}{\checked}	&	-	&	\cellcolor{green!45}{\checked}	&	-	&	\cellcolor{green!45}{\checked}	&	-	&	\cellcolor{green!45}{\checked}	&	-	&	\cellcolor{green!45}{\checked}	&	-	&	m22	&	\cellcolor{green!45}{\checked}	&	-	&	\cellcolor{green!45}{\checked}	&	-	&	\cellcolor{green!45}{\checked}	&	-	&	\cellcolor{green!45}{\checked}	&	-	&	\cellcolor{red!45}{$\times$}	&	1	&	\cellcolor{green!45}{\checked}	&	-	&	\cellcolor{green!45}{\checked}	&	-	&	c22	&	\cellcolor{red!45}{$\times$}	&	1	&	\cellcolor{red!45}{$\times$}	&	-	&	\cellcolor{red!45}{$\times$}	&	1	&	\cellcolor{red!45}{$\times$}	&	1	&	\cellcolor{green!45}{\checked}	&	-	&	\cellcolor{red!45}{$\times$}	&	1	&	\cellcolor{green!45}{\checked}	&	-
		\\ \rule{0pt}{8pt} 																																																																																								
		s23	&	\cellcolor{green!45}{\checked}	&	-	&	\cellcolor{red!45}{$\times$}	&	-	&	\cellcolor{green!45}{\checked}	&	-	&	\cellcolor{green!45}{\checked}	&	-	&	\cellcolor{green!45}{\checked}	&	-	&	\cellcolor{green!45}{\checked}	&	-	&	\cellcolor{green!45}{\checked}	&	-	&	m23	&	\cellcolor{green!45}{\checked}	&	-	&	\cellcolor{green!45}{\checked}	&	-	&	\cellcolor{green!45}{\checked}	&	-	&	\cellcolor{green!45}{\checked}	&	-	&	\cellcolor{red!45}{$\times$}	&	1	&	\cellcolor{red!45}{$\times$}	&	1	&	\cellcolor{green!45}{\checked}	&	-	&	c23	&	\cellcolor{red!45}{$\times$}	&	-	&	\cellcolor{red!45}{$\times$}	&	-	&	\cellcolor{green!45}{\checked}	&	-	&	\cellcolor{green!45}{\checked}	&	-	&	\cellcolor{green!45}{\checked}	&	-	&	\cellcolor{green!45}{\checked}	&	-	&	\cellcolor{red!45}{$\times$}	&	-
		\\ \rule{0pt}{8pt} 																																																																																								
		s24	&	\cellcolor{green!45}{\checked}	&	-	&	\cellcolor{red!45}{$\times$}	&	-	&	\cellcolor{green!45}{\checked}	&	-	&	\cellcolor{green!45}{\checked}	&	-	&	\cellcolor{green!45}{\checked}	&	-	&	\cellcolor{green!45}{\checked}	&	-	&	\cellcolor{green!45}{\checked}	&	-	&	m24	&	\cellcolor{green!45}{\checked}	&	-	&	\cellcolor{green!45}{\checked}	&	-	&	\cellcolor{green!45}{\checked}	&	-	&	\cellcolor{green!45}{\checked}	&	-	&	\cellcolor{red!45}{$\times$}	&	1	&	\cellcolor{green!45}{\checked}	&	-	&	\cellcolor{green!45}{\checked}	&	-	&	c24	&	\cellcolor{red!45}{$\times$}	&	-	&	\cellcolor{red!45}{$\times$}	&	-	&	\cellcolor{green!45}{\checked}	&	-	&	\cellcolor{green!45}{\checked}	&	-	&	\cellcolor{red!45}{$\times$}	&	1	&	\cellcolor{green!45}{\checked}	&	-	&	\cellcolor{green!45}{\checked}	&	-
		\\ \rule{0pt}{8pt} 																																																																																								
		s25	&	\cellcolor{green!45}{\checked}	&	-	&	\cellcolor{green!45}{\checked}	&	2	&	\cellcolor{green!45}{\checked}	&	-	&	\cellcolor{green!45}{\checked}	&	-	&	\cellcolor{green!45}{\checked}	&	-	&	\cellcolor{green!45}{\checked}	&	-	&	\cellcolor{green!45}{\checked}	&	-	&	m25	&	\cellcolor{green!45}{\checked}	&	-	&	\cellcolor{green!45}{\checked}	&	-	&	\cellcolor{green!45}{\checked}	&	-	&	\cellcolor{green!45}{\checked}	&	-	&	\cellcolor{red!45}{$\times$}	&	1	&	\cellcolor{green!45}{\checked}	&	-	&	\cellcolor{green!45}{\checked}	&	-	&	c25	&	\cellcolor{red!45}{$\times$}	&	-	&	\cellcolor{red!45}{$\times$}	&	2	&	\cellcolor{red!45}{$\times$}	&	1	&	\cellcolor{red!45}{$\times$}	&	1	&	\cellcolor{red!45}{$\times$}	&	1	&	\cellcolor{green!45}{\checked}	&	-	&	\cellcolor{red!45}{$\times$}	&	1
		\\ \rule{0pt}{8pt} 																																																																																								
		s26	&	\cellcolor{green!45}{\checked}	&	-	&	\cellcolor{green!45}{\checked}	&	-	&	\cellcolor{green!45}{\checked}	&	-	&	\cellcolor{green!45}{\checked}	&	-	&	\cellcolor{green!45}{\checked}	&	-	&	\cellcolor{red!45}{$\times$}	&	1	&	\cellcolor{green!45}{\checked}	&	-	&	m26	&	\cellcolor{green!45}{\checked}	&	-	&	\cellcolor{green!45}{\checked}	&	-	&	\cellcolor{green!45}{\checked}	&	-	&	\cellcolor{green!45}{\checked}	&	-	&	\cellcolor{green!45}{\checked}	&	-	&	\cellcolor{green!45}{\checked}	&	-	&	\cellcolor{green!45}{\checked}	&	-	&	c26	&	\cellcolor{green!45}{\checked}	&	-	&	\cellcolor{red!45}{$\times$}	&	-	&	\cellcolor{green!45}{\checked}	&	-	&	\cellcolor{green!45}{\checked}	&	-	&	\cellcolor{green!45}{\checked}	&	-	&	\cellcolor{green!45}{\checked}	&	-	&	\cellcolor{green!45}{\checked}	&	-
		\\ \rule{0pt}{8pt} 																																																																																								
		s27	&	\cellcolor{green!45}{\checked}	&	-	&	\cellcolor{green!45}{\checked}	&	1	&	\cellcolor{green!45}{\checked}	&	-	&	\cellcolor{green!45}{\checked}	&	-	&	\cellcolor{green!45}{\checked}	&	-	&	\cellcolor{green!45}{\checked}	&	-	&	\cellcolor{green!45}{\checked}	&	-	&	m27	&	\cellcolor{green!45}{\checked}	&	1	&	\cellcolor{green!45}{\checked}	&	1	&	\cellcolor{red!45}{$\times$}	&	1	&	\cellcolor{red!45}{$\times$}	&	1	&	\cellcolor{red!45}{$\times$}	&	1	&	\cellcolor{green!45}{\checked}	&	-	&	\cellcolor{green!45}{\checked}	&	-	&	c27	&	\cellcolor{green!45}{\checked}	&	1	&	\cellcolor{green!45}{\checked}	&	-	&	\cellcolor{green!45}{\checked}	&	-	&	\cellcolor{green!45}{\checked}	&	-	&	\cellcolor{green!45}{\checked}	&	-	&	\cellcolor{green!45}{\checked}	&	-	&	\cellcolor{green!45}{\checked}	&	-
		\\ \rule{0pt}{8pt} 																																																																																								
		s28	&	\cellcolor{red!45}{$\times$}	&	-	&	\cellcolor{green!45}{\checked}	&	-	&	\cellcolor{green!45}{\checked}	&	-	&	\cellcolor{green!45}{\checked}	&	-	&	\cellcolor{green!45}{\checked}	&	-	&	\cellcolor{green!45}{\checked}	&	-	&	\cellcolor{green!45}{\checked}	&	-	&	m28	&	\cellcolor{green!45}{\checked}	&	-	&	\cellcolor{green!45}{\checked}	&	-	&	\cellcolor{green!45}{\checked}	&	-	&	\cellcolor{green!45}{\checked}	&	-	&	\cellcolor{red!45}{$\times$}	&	1	&	\cellcolor{green!45}{\checked}	&	-	&	\cellcolor{green!45}{\checked}	&	-	&	c28	&	\cellcolor{green!45}{\checked}	&	2	&	\cellcolor{green!45}{\checked}	&	1	&	\cellcolor{red!45}{$\times$}	&	1	&	\cellcolor{red!45}{$\times$}	&	1	&	\cellcolor{red!45}{$\times$}	&	1	&	\cellcolor{red!45}{$\times$}	&	1	&	\cellcolor{red!45}{$\times$}	&	1
		\\ \rule{0pt}{8pt} 																																																																																								
		s29	&	\cellcolor{green!45}{\checked}	&	-	&	\cellcolor{red!45}{$\times$}	&	-	&	\cellcolor{green!45}{\checked}	&	-	&	\cellcolor{green!45}{\checked}	&	-	&	\cellcolor{green!45}{\checked}	&	-	&	\cellcolor{green!45}{\checked}	&	-	&	\cellcolor{green!45}{\checked}	&	-	&	m29	&	\cellcolor{red!45}{$\times$}	&	-	&	\cellcolor{red!45}{$\times$}	&	-	&	\cellcolor{green!45}{\checked}	&	-	&	\cellcolor{green!45}{\checked}	&	-	&	\cellcolor{green!45}{\checked}	&	-	&	\cellcolor{green!45}{\checked}	&	-	&	\cellcolor{red!45}{$\times$}	&	1	&	c29	&	\cellcolor{red!45}{$\times$}	&	3	&	\cellcolor{red!45}{$\times$}	&	4	&	\cellcolor{red!45}{$\times$}	&	1	&	\cellcolor{red!45}{$\times$}	&	1	&	\cellcolor{red!45}{$\times$}	&	1	&	\cellcolor{red!45}{$\times$}	&	1	&	\cellcolor{red!45}{$\times$}	&	-
		\\ \rule{0pt}{8pt} 																																																																																								
		s30	&	\cellcolor{red!45}{$\times$}	&	-	&	\cellcolor{red!45}{$\times$}	&	-	&	\cellcolor{green!45}{\checked}	&	-	&	\cellcolor{green!45}{\checked}	&	-	&	\cellcolor{green!45}{\checked}	&	-	&	\cellcolor{red!45}{$\times$}	&	1	&	\cellcolor{red!45}{$\times$}	&	1	&	m30	&	\cellcolor{red!45}{$\times$}	&	-	&	\cellcolor{red!45}{$\times$}	&	-	&	\cellcolor{red!45}{$\times$}	&	1	&	\cellcolor{green!45}{\checked}	&	-	&	\cellcolor{red!45}{$\times$}	&	1	&	\cellcolor{red!45}{$\times$}	&	1	&	\cellcolor{red!45}{$\times$}	&	-	&	c30	&	\cellcolor{green!45}{\checked}	&	3	&	\cellcolor{green!45}{\checked}	&	1	&	\cellcolor{red!45}{$\times$}	&	1	&	\cellcolor{green!45}{\checked}	&	-	&	\cellcolor{green!45}{\checked}	&	-	&	\cellcolor{green!45}{\checked}	&	-	&	\cellcolor{red!45}{$\times$}	&	1 \\
		\bottomrule
	\end{tabular}
\end{table*}

\subsection{[RQ1]: What are the differences in detection performance between LLMs and EDA tools?}

We utilize the LintLLM framework to enhance LLMs, aiming to evaluate its state-of-the-art performance in defect detection. The detailed experimental results are shown in TABLE \ref{Detail_Results}. 

Simple benchmark. The performance of LLMs and commercial EDA tool is comparable. However, Verilator failed to detect 6 defects and generated 5 false positives.

Medium benchmark. The advantages of LLMs are highlighted. DeepSeek V2.5 (\ding{185}) and o1-mini (\ding{188}) have only one (m27) or two (m29, m30) defects not detected, respectively, and only one false positive. However, Llama-3.1 (\ding{184}) and GPT-4 (\ding{186}) perform poorly in complex DUTs.

Complex benchmark. Although both LLMs and EDA tools face challenges, LLMs still outperform EDA tools. o1-mini (\ding{188}) correctly detects 18 defects and generates 9 false positives, performing best among the evaluated models. In contrast, commercial EDA tool (\ding{182}) only correctly detects 12 defects and generates 16 false positives. Meanwhile, Verilator (\ding{183}) produced 14 false positives and only correctly detected 12 defects out of 30 complex DUTs. 

\begin{figure}[htbp]
	\centering
		\begin{subfigure}[]{0.24\textwidth}
				\includegraphics[width=\textwidth]{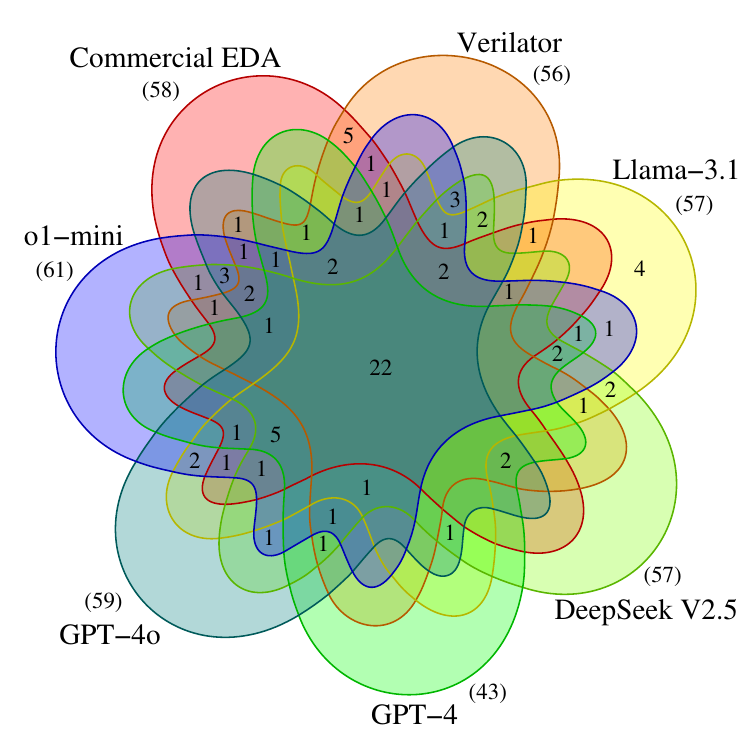}\vspace{-2mm}
				\caption{Correct: Original}
				\label{original_c}
			\end{subfigure}
		\begin{subfigure}[]{0.24\textwidth}
				\includegraphics[width=\textwidth]{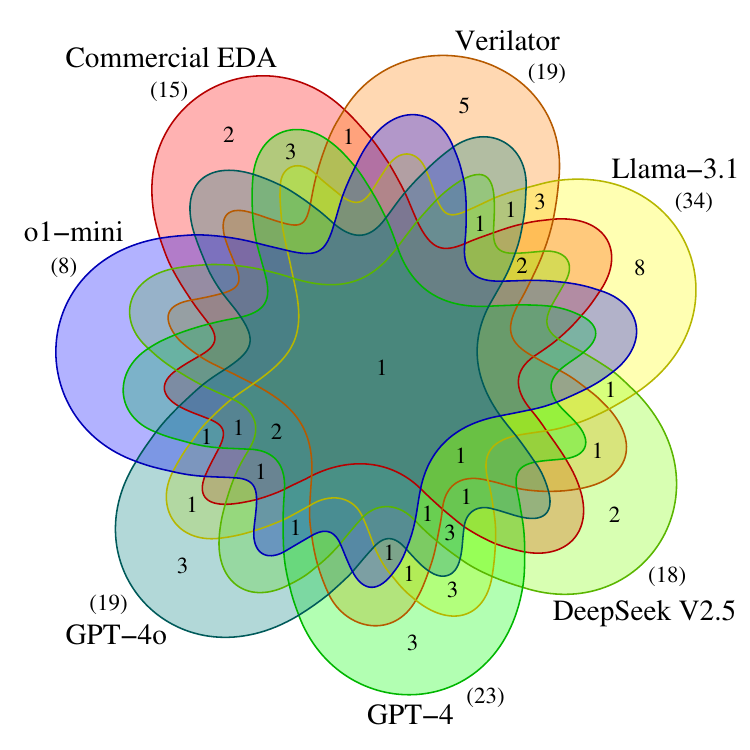}\vspace{-2mm}
				\caption{False Positive: Original}
				\label{original_f}
			\end{subfigure}
	\begin{subfigure}[]{0.24\textwidth}
		\includegraphics[width=\textwidth]{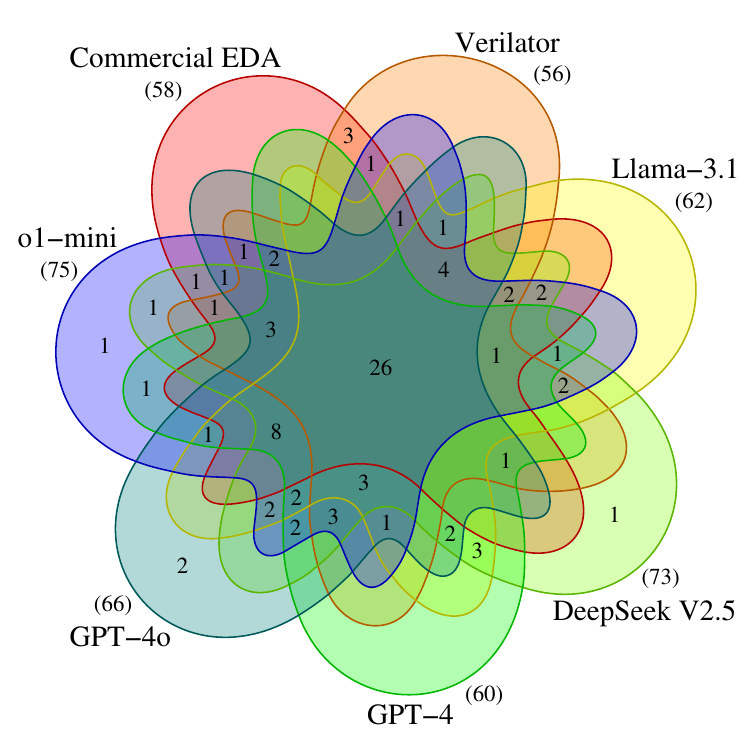}\vspace{-2mm}
		\caption{Correct: LintLLM}
		\label{Both_methods_c}
	\end{subfigure}
	\begin{subfigure}[]{0.24\textwidth}
		\includegraphics[width=\textwidth]{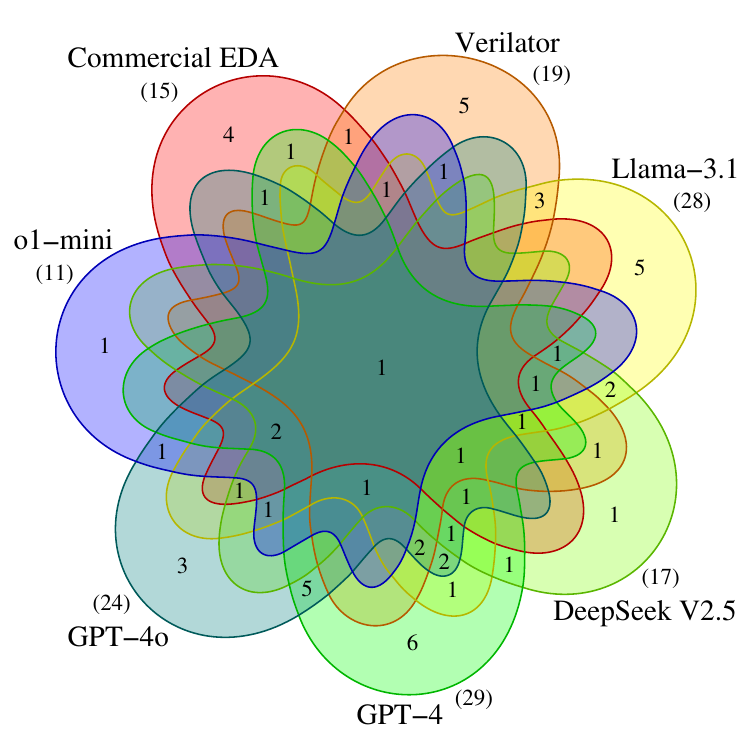}\vspace{-2mm}
		\caption{False Positive: LintLLM}
		\label{Both_methods_f}
	\end{subfigure}
	\caption{Comparison of defect detection results between the original model and LintLLM. The numbers in the overlapping area indicate the number of DUTs detected as correct or false positive by multiple models. It shows that LintLLM outperforms the Original method.}
	\label{overview_result}
	
\end{figure}

The detection results of the Original models and the LintLLM optimization are shown in Fig. \ref{overview_result}. All LLMs show improvements in the LintLLM scenario. In terms of false positives, Llama-3.1 and DeepSeek V2.5 reduce 6 and 1 false positives, respectively. However, the false positives of the GPT series models (GPT-4, GPT-4o, and o1-mini) increase slightly, which will be discussed in RQ2 (Section \ref{rq2}).

\textbf{Summary}: LLMs outperform EDA tools in both detection accuracy and false positive control. Among 90 DUTs, o1-mini enhanced by LintLLM detects 75 defects and produces 11 false positives, while commercial EDA tool only detects 58 defects and produces 15 false positives. It demonstrates the powerful potential of LLMs in code defect detection.

\begin{figure*}[htbp]
	\centering
	\begin{subfigure}[b]{0.195\textwidth}
		\includegraphics[width=\textwidth]{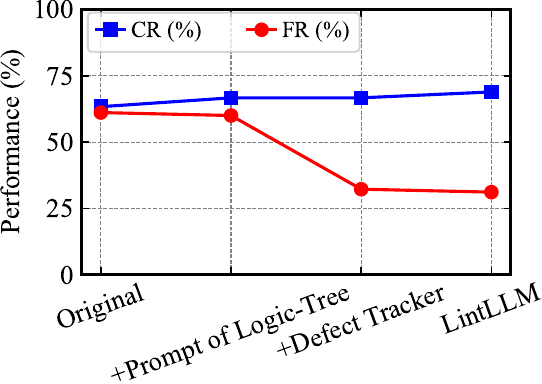} \vspace{-6mm}
		\caption{Llama-3.1}
	\end{subfigure}
	\begin{subfigure}[b]{0.195\textwidth}
		\includegraphics[width=\textwidth]{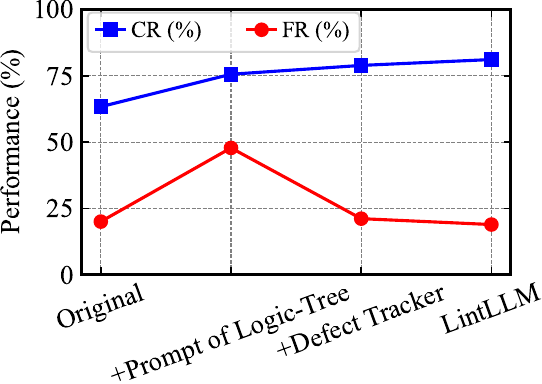} \vspace{-6mm}
		\caption{DeepSeek V2.5}
	\end{subfigure}
	\begin{subfigure}[b]{0.195\textwidth}
		\includegraphics[width=\textwidth]{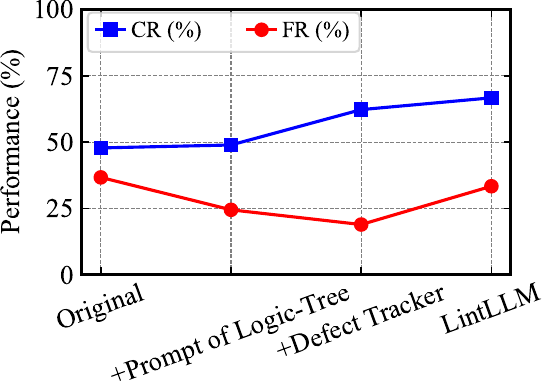} \vspace{-6mm}
		\caption{GPT-4}
	\end{subfigure}
	\begin{subfigure}[b]{0.195\textwidth}
		\includegraphics[width=\textwidth]{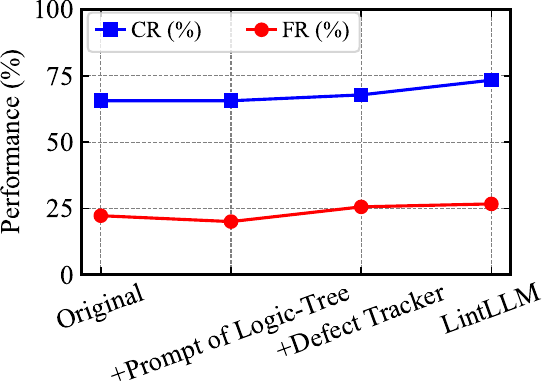} \vspace{-6mm}
		\caption{GPT-4o}
	\end{subfigure}
	\begin{subfigure}[b]{0.195\textwidth}
		\includegraphics[width=\textwidth]{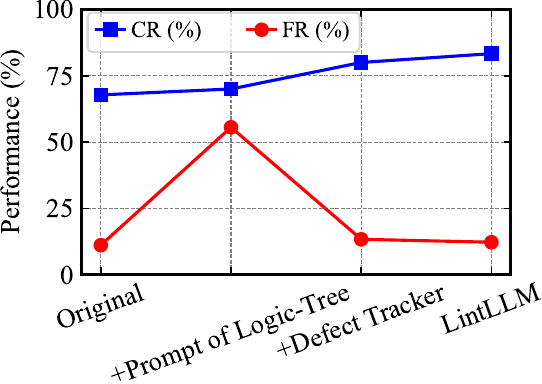} \vspace{-6mm}
		\caption{o1-mini}
	\end{subfigure}
	\caption{The impact of different prompting methods on defect detection. The CR is better when higher (\textcolor{blue}{$\nearrow$}), and FR is better when lower (\textcolor{red}{$\searrow$}). It is clearly shown that Prompt of Logic-Tree improves the correct rate, and Defect Tracker reduces the false-positive rate. With the combined effect of the two methods, LLMs have the best performance.}
	\label{different_methods}
\end{figure*}

\subsection{[RQ2]: Why our methods can improve the detection performance of LLM? \label{rq2}}

To explore the performance of different methods, we conducted experiments using Prompt of Logic-Tree and Defect Tracker, respectively. The results are shown in Fig. \ref{different_methods}. Prompt of Logic-Tree improves correct rate for GPT-4 from 47.78\% to 48.89\%, while the false-positive rate drops to 24.44\%. This shows that the method can help LLMs make more accurate judgments on defect detection. Although the correct rate of DeepSeek V2.5 significantly increases to 75.56\%, the false-positive rate also increases to 47.78\%. This is because the Prompt of Logic-Tree mainly enhances LLMs to follow the defect detection process, but ignores the case where multiple defects exist in the code.

Defect Tracker reduces false-positive rates by tracking multiple defect paths in the code and locating the main defect that causes multiple defects. Except for GPT-4o, the false-positive rates of each model show a significant reduction, while improving the detection correctness. These results indicate that the Defect Tracker is very effective in reducing false-positive rate, especially for models with high initial false-positive rates, such as Llama-3.1 and GPT-4. 

Combining the two methods improves the model's overall performance, yet the false-positive rate of GPT series models (GPT-4, GPT-4o, and o1-mini) does not drop as expected. We speculate two main reasons. (\rmnum{1}) Possibly some undisclosed security mechanisms prevent the knowledge structure of the model from being changed by prompts. (\rmnum{2}) Combining the two methods may produce information conflicts, leading to an increase in false-positive rates. Future research will focus on dynamically adjusting the prompts according to the characteristics of LLMs.


\textbf{Summary}: Prompt of Logic-Tree uses logical prompts to guide LLMs for gradually detecting defects and improving the accuracy, such as the accuracy of DeepSeek V2.5 increasing from 63.33\% to 75.56\%. Defect Tracker effectively reduces the false-positive rate by locating the main defects, with o1-mini's false-positive rate dropping by 42.23\%. The comprehensive performance is improved after combining both methods.

\subsection{[RQ3]: What are the advantages of LLMs in detection costs compared with commercial EDA tools?}

To comprehensively explore the cost advantages of LLMs, we conducted further analysis. Based on provider pricing of LLMs \cite{price}, the cost of processing 1 million tokens (about 80,000 lines of code) is \$3 for inputting and \$12 for outputting. The input tokens include DUTs and prompts, and the output tokens cover the analysis process and generated defect reports. After conversion, the total cost of detecting 80,000 lines of code is about \$20. For commercial EDA tools, we assume the annual license fee is about \$1.2 million.

Firstly, we analyze the relationship between code volume and cost, as shown in Fig. 6(a). When the annual code volume is less than 4800 million lines (equivalent to the code volume of a development team of 100,000 people in one year), using LLMs is more economical (Region I). However, when this threshold is exceeded, the cost advantage of using commercial EDA tool is more obvious (Region II). Therefore, LLMs is a more cost-effective solution for small and medium-sized enterprises and research institutions with less code volume.

Secondly, we further quantified the cost through a real case. A medium-sized CPU control unit contains about 1,000 lines of Verilog code, which meets the context tokens restrictions of LLMs. The cost of using LLMs to perform 1 defect detection is \$0.285, and it runs 1,000 times per day, with an annual cost of about \$0.1 million. Fig. 6(b) shows the relationship between cost and detection performance. Based on commercial EDA tool, investing \$0.1 million to integrate LLMs will improve performance by approximately 20.00\%, achieving a correct rate of 87.78\% and a false-positive rate of 7.78\%.

\begin{figure}[htbp]
	\centering
	\begin{subfigure}[]{0.24\textwidth}
		\includegraphics[width=\textwidth]{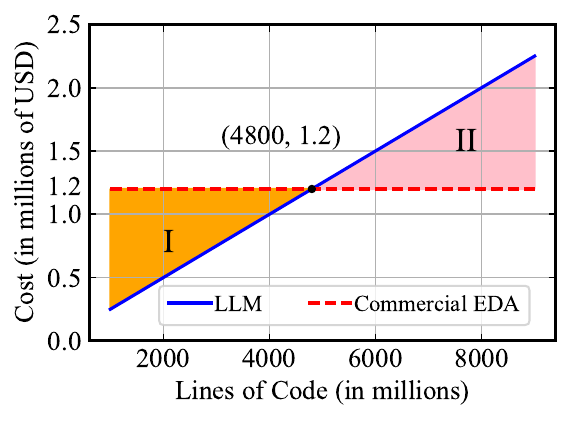}
		\caption{Lines of Code vs. Cost}
		\label{cost1}
	\end{subfigure}
	\begin{subfigure}[]{0.24\textwidth}
		\includegraphics[width=\textwidth]{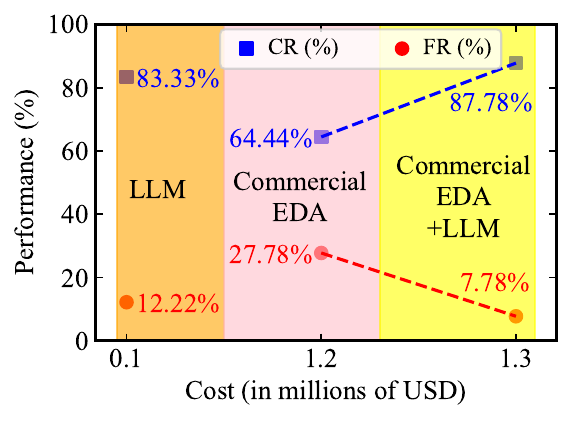}
		\caption{Cost vs. Performance}
		\label{cost2}
	\end{subfigure}
	\caption{Cost Comparison of commercial EDA Tool and LLM. LLM achieves better performance at one-tenth the cost of EDA tools.}
	\label{cost}
\end{figure}

\textbf{Summary}: For small and medium-sized enterprises and research institutions processing fewer than 4,800 million lines of code annually, LLM offers significant advantages in reducing defect detection costs. Additionally, integrating LLM with commercial EDA tool at a lower cost will improve performance by approximately 20\%.

\section{Conclusion}

This paper introduces LintLLM, an open-source Linting framework that leverages LLMs to detect defects in Verilog code. This study demonstrates that LLM is an efficient and economical solution for hardware defect detection. In the future, we will explore using LLMs to automatically repair defects after completing detection and finally obtain high-quality RTL designs. Additionally, integrating LintLLM into  IDEs is worth considering.
\clearpage

\bibliographystyle{IEEEtran} 
{\footnotesize \bibliography{Mybib.bib}}
\nocite{*}

\end{document}